\documentclass[%
 reprint,
 superscriptaddress,
 showpacs, 
 amsmath,amssymb,
 aps,
 prl,
 floatfix,
]{revtex4-1}

\usepackage[pdftex]{graphicx}
    \graphicspath{{figures/}}
    \DeclareGraphicsExtensions{.pdf,.png}
\usepackage{dcolumn}
\usepackage{bm}
\usepackage{color}
\usepackage{multirow}

\begin{document}

\title{Trilayer dusty plasma lattice: structure and dynamics}

\author{Hong Pan}
\affiliation{Department of Physics, Boston College, Chestnut Hill, Massachusetts, 02467, USA}

\author{Gabor J. Kalman}
\affiliation{Department of Physics, Boston College, Chestnut Hill, Massachusetts, 02467, USA}

\author{Peter Hartmann}
\affiliation{Institute for Solid State Physics and Optics, Wigner Research Centre, Hungarian Academy of Sciences, P.O.Box. 49, H-1525 Budapest, Hungary}
\affiliation{Department of Physics, Boston College, Chestnut Hill, Massachusetts, 02467, USA}

\author{Zolt\'an Donk\'o}
\affiliation{Institute for Solid State Physics and Optics, Wigner Research Centre, Hungarian Academy of Sciences, P.O.Box. 49, H-1525 Budapest, Hungary}
\affiliation{Department of Physics, Boston College, Chestnut Hill, Massachusetts, 02467, USA}

\date{\today}

\begin{abstract}
Abstract:
In this paper, we studied the structure and dynamics for a trilayer Yukawa crystal. We firstly studied the optimal lattice structure by comparing lattice interaction energy from different lattice distribution, after that, we did the collective mode calculation and analyzed its eigenvectors, if the lattice structure is stable, all the eigenvalues of the dynamical matrix should be positive.
\end{abstract}

\pacs{52.27.Gr, 52.27.Lw, 63.20.-e}
\maketitle

Wigner proposed that the electron can form crystal at low electron density \cite{Wigner1934}. Wigner crystal can be created by electrons in liquid helium surface, dusty plasma or colloidal particles. If the the neutralizing background is fixed, it is Coulomb plasma or one component plasma(OCP), if the background is polarizable, due to the Debye screening, the interaction between charged particles becomes Yukawa potential. Because it is easier to do measurement in experiment, people often study 2D system. Bonsall\cite{Lynn1977} firstly studied the static and dynamical properties of a 2D Coulomb Wigner crystal. Gann's Monte Carlo simulation result\cite{Gann1979} and Thomas's experimental result\cite{Thomas1994} confirmed the stable 2D lattice structure is hexagonal. Dispersion relation in 2D Yukawa lattice was firstly calculated by Peeters\cite{Peeters1987}in 1987, followed by works on Yukawa system\cite{Thomas2006,Peter2007,Donko2008,Morfill2003,Hou2009,Nunomura2002,Wang2001}. Normally, a 2D system needs a confining potential in the direction perpendicular to the 2D plane. The dusty plasma system under weak confining may form multilayer structure\cite{Klumov2008,Teng2003}. The lattice structure transition with different confining strength was studied in works\cite{Lowen2009a,Lowen2009b,Pieranski1983,Schmidt1996,Ramiro2007,Fontecha2005}. Bilayer Coulomb lattice system was firstly studied by Peeters\cite{Goldoni1995}, followed by some works on bilayer Yukawa system\cite{Rene2003,Rene2006}. In this letter, we try to study a trilayer lattice system, each layer is a perfect lattice and particles have only inplane movement.

We assume each layer has the same kind of lattice structure, due to the symmetry, the unit cell is either rhombic or rectangle, here we can tune two parameters: rhombic angle and the aspect ratio. Originally, we put both the top and bottom layers at the geometric center of the middle layer(unshifted), then we can shift the top and bottom layer along a symmetric axis in opposite directions, hence we have another tuned parameter, shift value. In rhombic case, suppose the primitive vectors for the middle layer are $\vec{a}=(xcos(\frac{\alpha}{2}),xsin(\frac{\alpha}{2}))$ and $\vec{b}=(-xcos(\frac{\alpha}{2}),xsin(\frac{\alpha}{2}))$, $x$ is the side length of a rhombic unit cell. In rectangle case, the primitive vectors for the middle layer are $\vec{a}=(x,0)$ and $\vec{b}=(0,tx)$, $x$ is the length of the horizontal side of the rectangle unit cell. If $\vec{r}$ is the the inplane vector of an arbitrary site in middle layer, then $\vec{r}+\frac{\vec{a}+\vec{b}}{2}+\vec{c}$ gives the position of a site in top layer, $\vec{r}+\frac{\vec{a}+\vec{b}}{2}-\vec{c}$ gives the position of a site in bottom layer, where $\vec{c}$ is the shift vector. In rhombic case, $\vec{c}=(0,s)$,  $0 \leqslant s\leqslant |\frac{\vec{a}+\vec{b}}{2}|$. In rectangle case, there are two kinds of shift: vertical shift and diagonal shift.The vertical shift is the same as that in rhombic case except that  $0 \leqslant s\leqslant \frac{|\vec{b}|}{2}$. In diagonal shift, $\vec{c}=s(1,t)$,  $0 \leqslant s\leqslant |\frac{\vec{a}+\vec{b}}{2}|$. If the shift vector $\vec{c}=0$, we call it unshifted lattice.

\begin{table}
\begin{tabular}{|c|c|c|c|}
 \hline
 type & $\vec{a}/x$ & $\vec{b}/x$ & $n_sx^2$ \\
 \hline
 rhombic  & $(cos(\frac{\alpha}{2}),sin(\frac{\alpha}{2}))$  & $(-cos(\frac{\alpha}{2}),sin(\frac{\alpha}{2}))$  &  $\frac{3}{sin(\alpha)}$\\
\hline
 rectangle  & $(1,0)$  & $(0,t)$  &  $\frac{3}{t}$\\
\hline
\end{tabular}
\caption{primitive vectors}
\label{table:1}
\end{table}

\begin{table}
\begin{tabular}{|c|c|c|}
 \hline
 middle & top & bottom \\
 \hline
 $0$  & $\frac{\vec{a}+\vec{b}}{2}+\vec{c}$   & $\frac{\vec{a}+\vec{b}}{2}-\vec{c}$   \\
\hline
\end{tabular}
\caption{layer shift}
\label{table:2}
\end{table}

\begin{table}
\begin{tabular}{|c|c|c|c|}
 \hline
 type & direction & $\vec{c}/s$  & s range \\
 \hline
 rhombic  & vertical   & (0,1) &   $0 \leqslant s\leqslant |\frac{\vec{a}+\vec{b}}{2}|$ \\
\hline
\multirow{2}{*}{rectangle} & vertical & (0,1) & $0 \leqslant s\leqslant \frac{|\vec{b}|}{2}$ \\
& diagonal & (1,t) &   $0 \leqslant s\leqslant |\frac{\vec{a}+\vec{b}}{2}|$ \\
\hline
\end{tabular}
\caption{shift type}
\label{table:3}
\end{table}

\begin{figure}[htbp]
\begin{center}
\includegraphics[width=0.4\columnwidth]{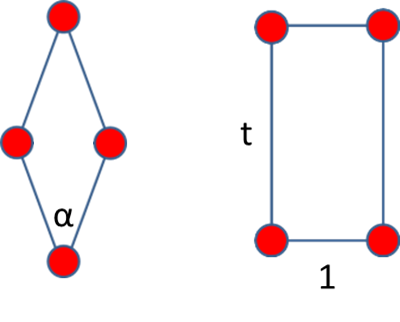}
\caption{Rhombic unit cell(left), rectangle unit cell(right),$\alpha$ is the rhombic angle, the ratio between the two sides of the rectangle is t
}
\label{fig:p1}
\end{center}
\end{figure}

We define the Wigner-Seitz(WS) radius based on $\pi a^2 n_s=1$,$a$ is the WS radius, $n_s$ is the projected surface density. In the following part of this paper, all the length variables are in the unit of $a$.

For a given lattice distribution, we calculate the Yukawa interaction energy from all the other neighbor sites within a given radius. The total energy is $E_{total}=3E_1+2E_2+E_3$, $E_1$ is the energy within each layer,$E_2$ is the energy between the nearest layers, $E_3$ is the energy between the top and bottom layers. 

\begin{figure}[htbp]
\begin{center}
\includegraphics[width=0.7\columnwidth]{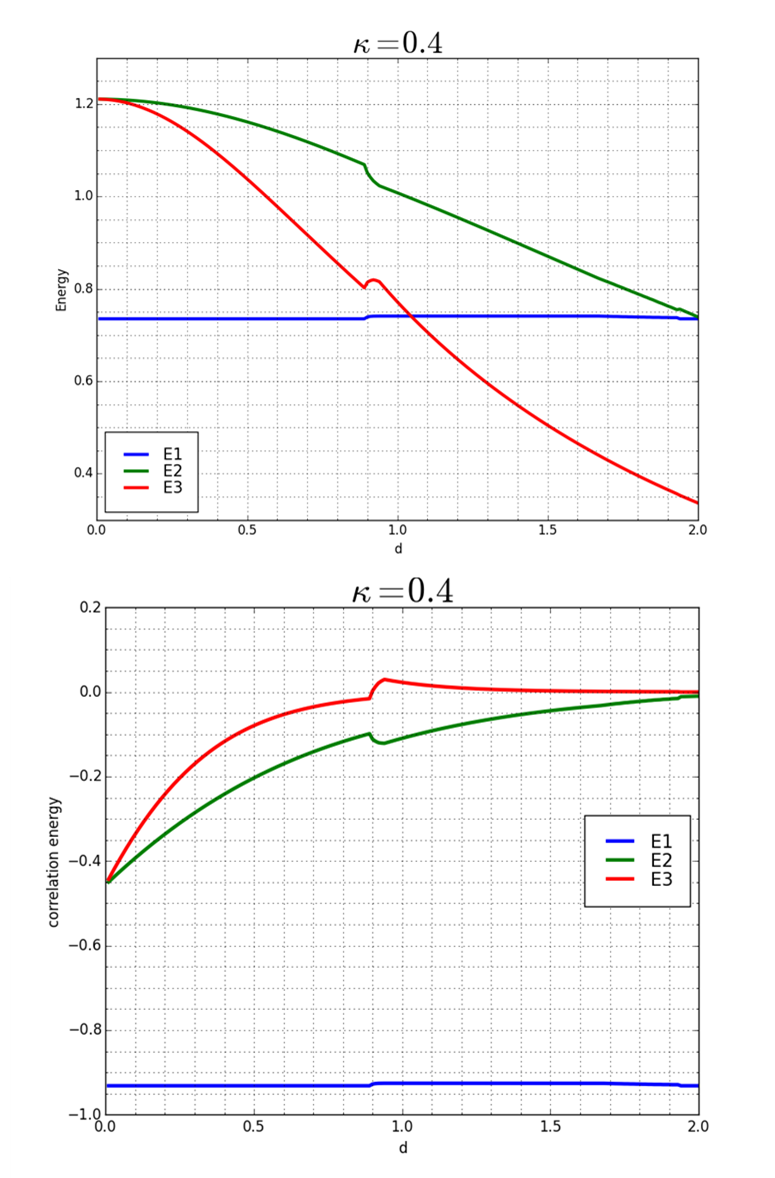}
\caption{Energy and correlation energy plot for $E_1,E_2,E_3$
}
\label{fig:e1}
\end{center}
\end{figure}

\begin{figure}[htbp]
\begin{center}
\includegraphics[width=\columnwidth]{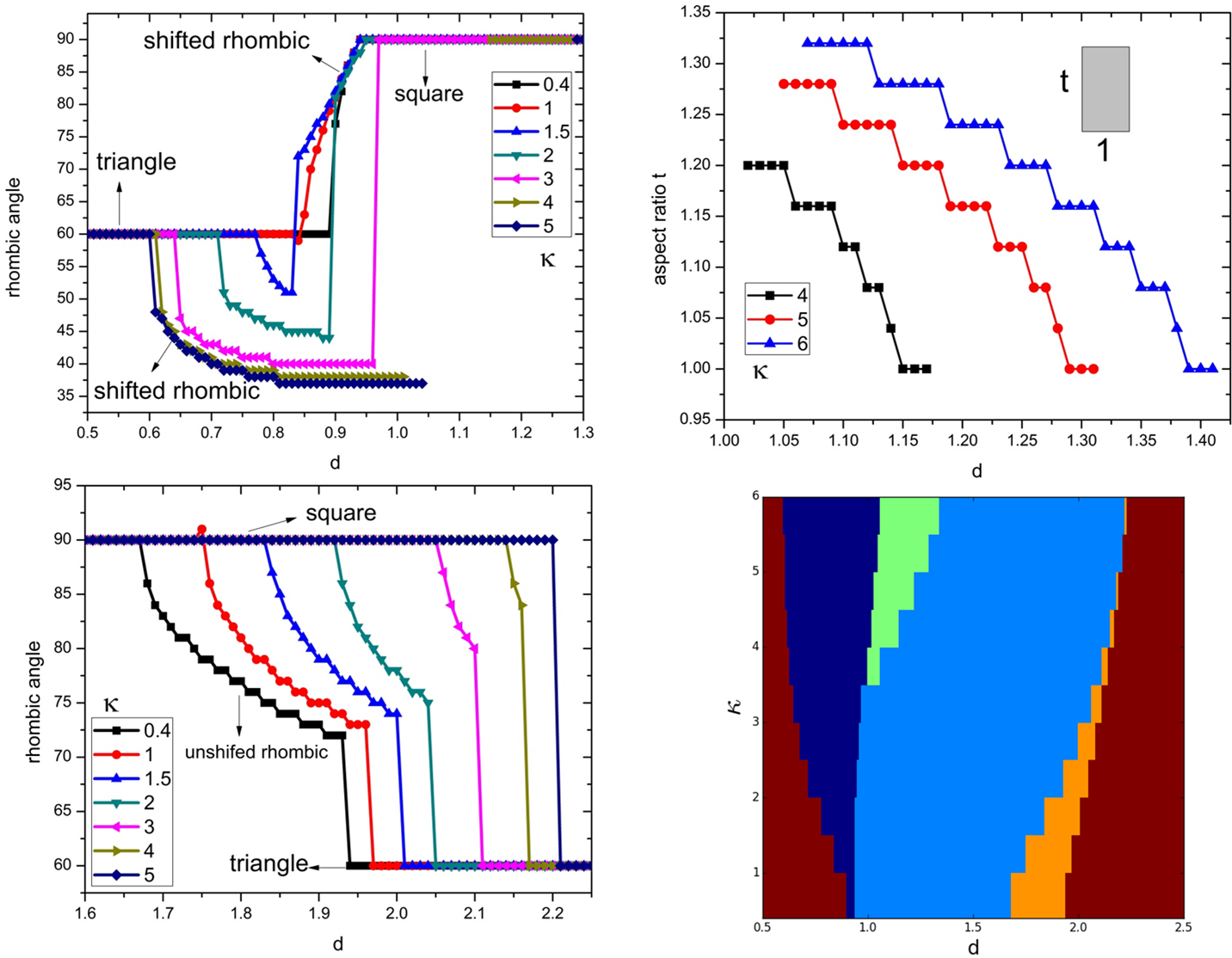}
\caption{phase diagram with different interlayer distance, the fourth image is the $\kappa-d$ phase diagram, the red is triangle lattice, the dark blue is the shifted rhombic, the green is rectangle, the light blue is square}
\label{fig:ps1}
\end{center}
\end{figure}

We set a value for the interlayer distance. Then compare the energy from different types of lattice, the one with the lowest energy is the optimal lattice. Fig \ref{fig:ps1} shows the phase as the interlayer distance and $\kappa$ change. Fig \ref{fig:e1} is the plot of energy for $\kappa=0.4$.

\begin{figure}[htbp]
\begin{center}
\includegraphics[width=\columnwidth]{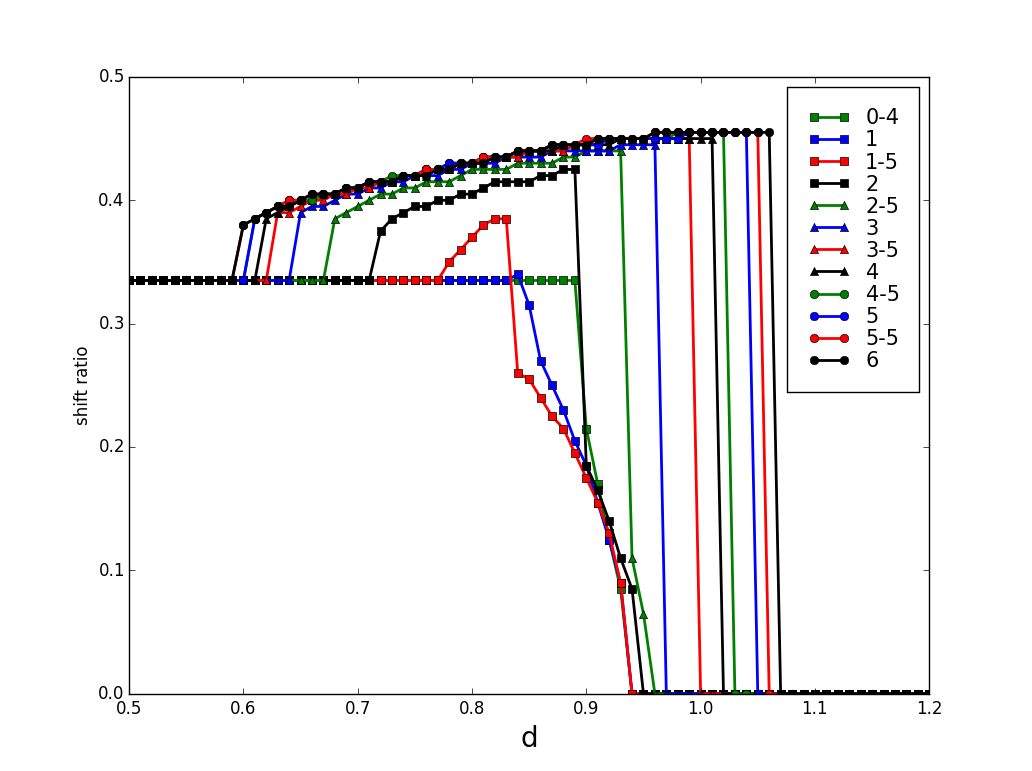}
\caption{The shift ratio with different d values, the shift ratio is the ratio between the shift value and the half diagonal side of the rhombus}
\label{fig:shift}
\end{center}
\end{figure}

\begin{figure}[htbp]
\begin{center}
\includegraphics[width=0.8\columnwidth]{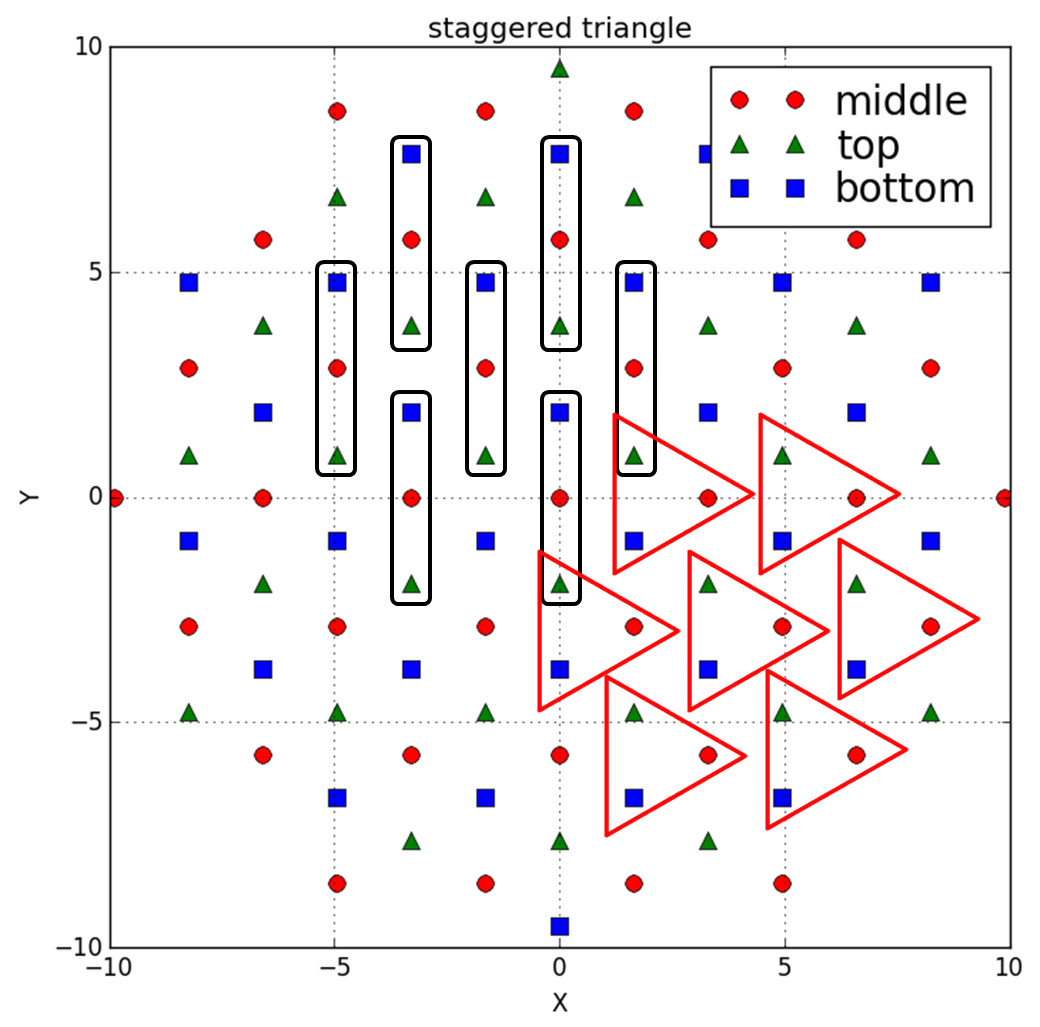}
\caption{top view of a staggered trilayer triangular lattice, the bar and triangle represents two possible unit cell options, the lattice structure of the new unit cell is the same as each layer}
\label{fig:p3}
\end{center}
\end{figure}

Based on the well-known result that the stable state of a single layer lattice is hexagonal, and a single layer hexagonal lattice can be decomposed into three hexagonal lattices, so when the interlayer distance is small, we should always get a staggered triangle lattice as in Fig \ref{fig:p3}. At very large interlayer distance, since the coupling between layers is very weak, we can treat each of them as a single layer, so it also must be a triangular lattice. In Fig \ref{fig:e1},it is the plot of energy $E_1,E_2,E_3$ as well as correlation energy. we can see there is kink at $d=0.9, 1.95$ because of the abrupt lattice phase transition. The correlation energy between top and bottom layer is positive at $d>0.9$, this indicates the lattice sites from the top layer are just right above the those from the bottom layer(unshifted lattice).

In the previous section, we have found the optimal lattice at a certain interlayer distance. But whether it is stable, we still don’t know, since our rhombic or rectangle unit cell can't map all the possible lattice distribution. the collective mode dispersion can provide useful evidence to check the stability. In this section, we will study the in-plane collective mode dispersion.

In order to obtain the collective mode dispersion, we need the dynamical matrix $M_{\mu\nu}^{ij}$, here, i, j is 1(middle),2(top),3(bottom); $\mu$, $\nu$ is x, y. So the dynamical matrix is a 6$\times$6 matrix.

\begin{equation} \label{eq:mat1}
M_{\mu\nu}^{ij}(\vec{k})=\delta_{ij}\sum\limits_{m=1,2,3} S_{\mu\nu}^{im}(0)-S_{\mu\nu}^{ij}(\vec{k})
\end{equation}

\begin{equation} \label{eq:mat2}
S_{\mu\nu}^{ij}(\vec{k})=\sum\limits_{n} \frac{\partial ^2 \varphi ^{ij} (\vec{r}_n)}{\partial\mu\partial\nu}e^{i\vec{k}\vec{r}_n}
\end{equation}

$\sum\limits_{n}$ means the lattice summation from all the sites within a given radius.The structure of dynamical matrix in the form 
$
\begin{pmatrix} 
M^{11} & M^{12} & M^{13} \\
M^{21} & M^{22} & M^{23} \\
M^{31} & M^{32} & M^{33}
\end{pmatrix} 
$
where each sub-block is 
$
\begin{pmatrix} 
M^{ij}_{xx} & M^{ij}_{xy}  \\
M^{ij}_{yx} & M^{ij}_{yy}  \\
\end{pmatrix} 
$
The dynamical matrix is always Hermitian. It always has real eigenvalues. If our lattice is stable, all the eigenvalues must be positive, or else, some eigenvalue in some k range may be negative. Here we show the dispersion result from three types of lattice: staggered triangle, overlapped square. Note: All the dispersion graphs are generated with very fine cut of k value, the step for k is 0.0002, so we can clearly see the what happens when two dispersion curves cross each other. 

Since the matrix has 6 dimensions, its eigen problem can only be solved by numerical calculation. Here we use the \textsl{Eigen C++} library to do the computation. When we display the six modes, we distinguish each mode by the continuity of eigenvector. We start to generate the eigenvalues from $k=0.0001$. Suppose $\lambda(k)$ is the eigenvalue, then $w(k)=\sqrt{\lambda(k)}$, if $\lambda(k)$ is negative, we set $w(k)=-\sqrt{-\lambda(k)}$. For given two consecutive wavevector values $k$ and $k+\Delta k$, we have six normalized eigenvectors $\vec{v}_i(k)$ and $\vec{v}_i(k+\Delta k)$ , $i=$1 to 6. We check the dot product value of $\vec{v}_i(k)\cdot \vec{v}_j(k+\Delta k)$, If its value has least deviation from 1, we display eigenvalue $w_i(k)$ and $w_j(k+\Delta)$ with the same color. As for the name in the legend, it will be discussed later.

\begin{figure}[htbp]
\begin{center}
\includegraphics[width=\columnwidth]{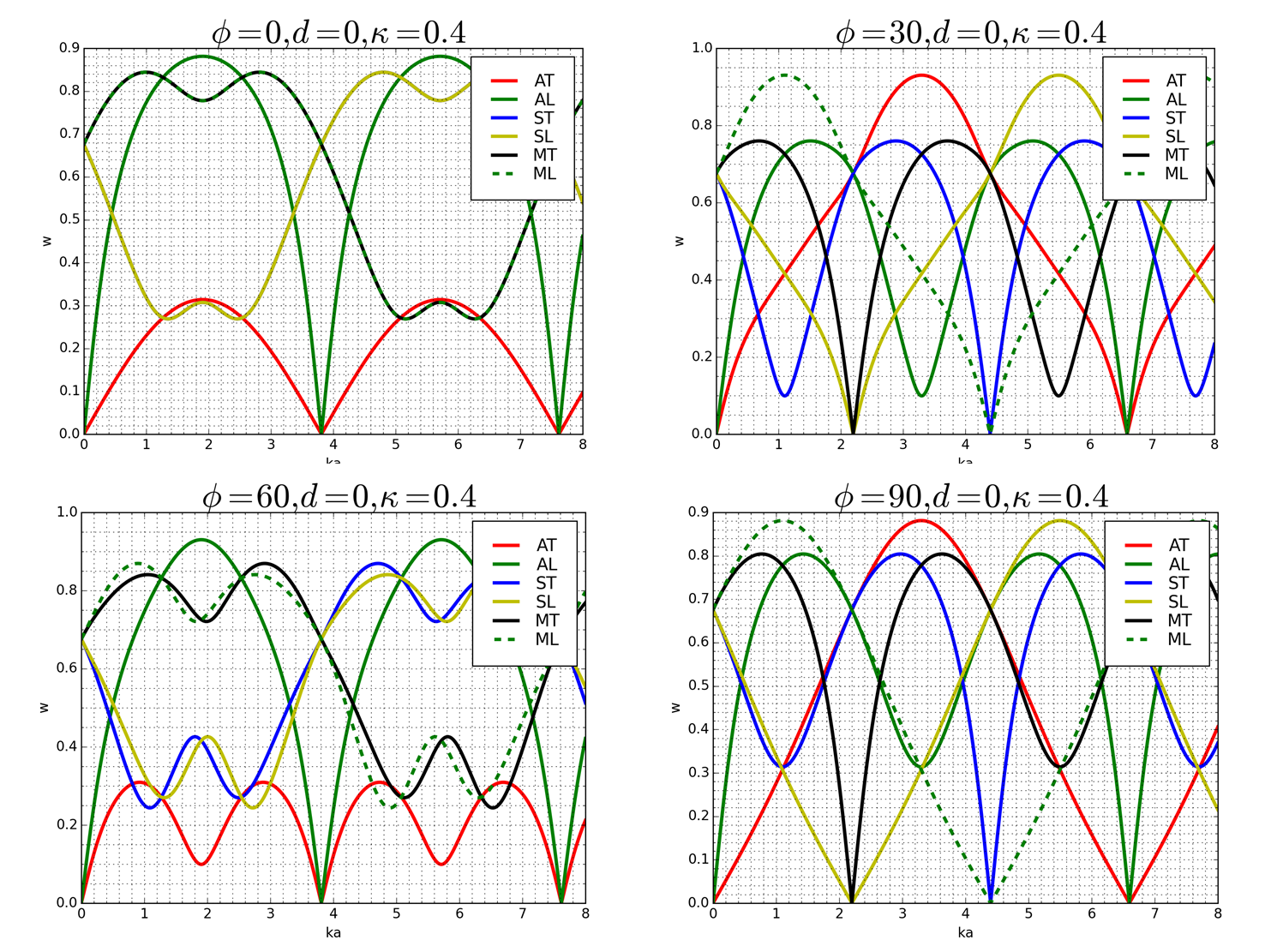}
\caption{dispersion curves when interlayer distance is zero}
\label{fig:dis1}
\end{center}
\end{figure}

\begin{figure}[htbp]
\begin{center}
\includegraphics[width=0.6\columnwidth]{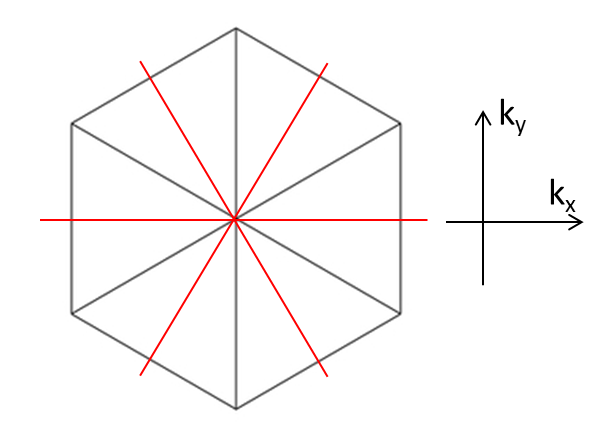}
\caption{Reciprocal lattice: red line is for the total lattice(ignore the label of different layer), black line is for each sublattice or the unit cell in Fig \ref{fig:p3}}
\label{fig:bz}
\end{center}
\end{figure}

In Fig \ref{fig:bz}, it is the reciprocal lattice, black lines are the axis of from each sublattice, red lines are the axis from the total single layer lattice(ignore the label difference). 

From Fig \ref{fig:dis1}, when k angle=30, 90, the dispersion curves are three sets of lines, they are exactly the same but with a shift along k axis, this is because the single triangle lattice contains three dilute triangle lattice. Along k angle=0, 30, the gapped mode is some new information which doesn't show up in single layer lattice. We think it is because of the lattice structure difference between the total single lattice and the new unit cell in Fig \ref{fig:p3}. The images from  angle 0 and 60 or 30 and 90 are almost same but still have some difference, we don't know the reason, perhaps because there is no 60 degree rotation symmetry for each unit cell. Since when d=0, the trilayer system is the same as a single hexagonal layer, recall the single layer lattice dispersion\cite{Thomas2006,Donko2008,Peter2007}, we do see it in Fig \ref{fig:dis1}.

\begin{figure}[htbp]
\begin{center}
\includegraphics[width=\columnwidth]{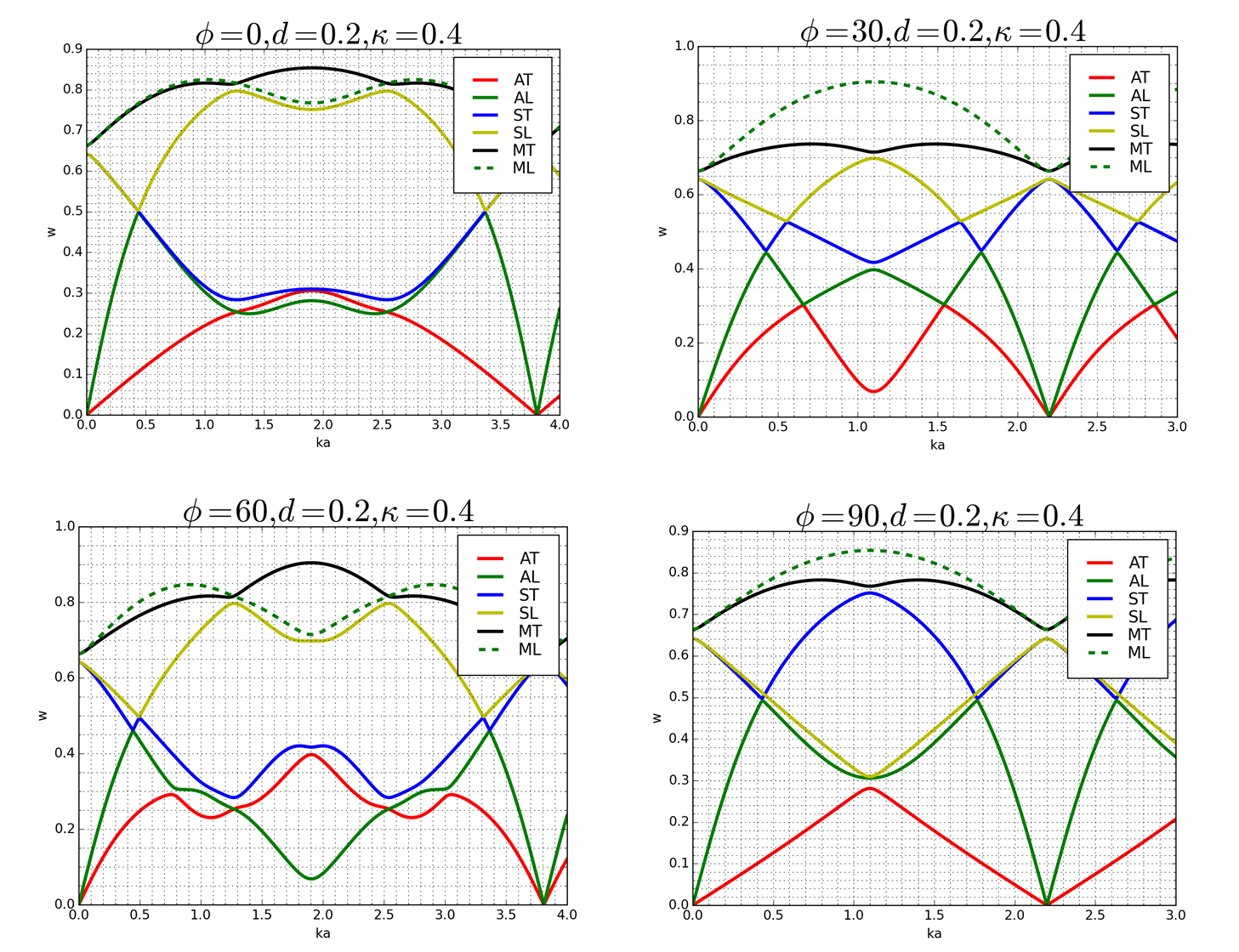}
\caption{dispersion curves for $d=0.2a$}
\label{fig:dis2}
\end{center}
\end{figure}

\begin{figure}[htbp]
\begin{center}
\includegraphics[width=\columnwidth]{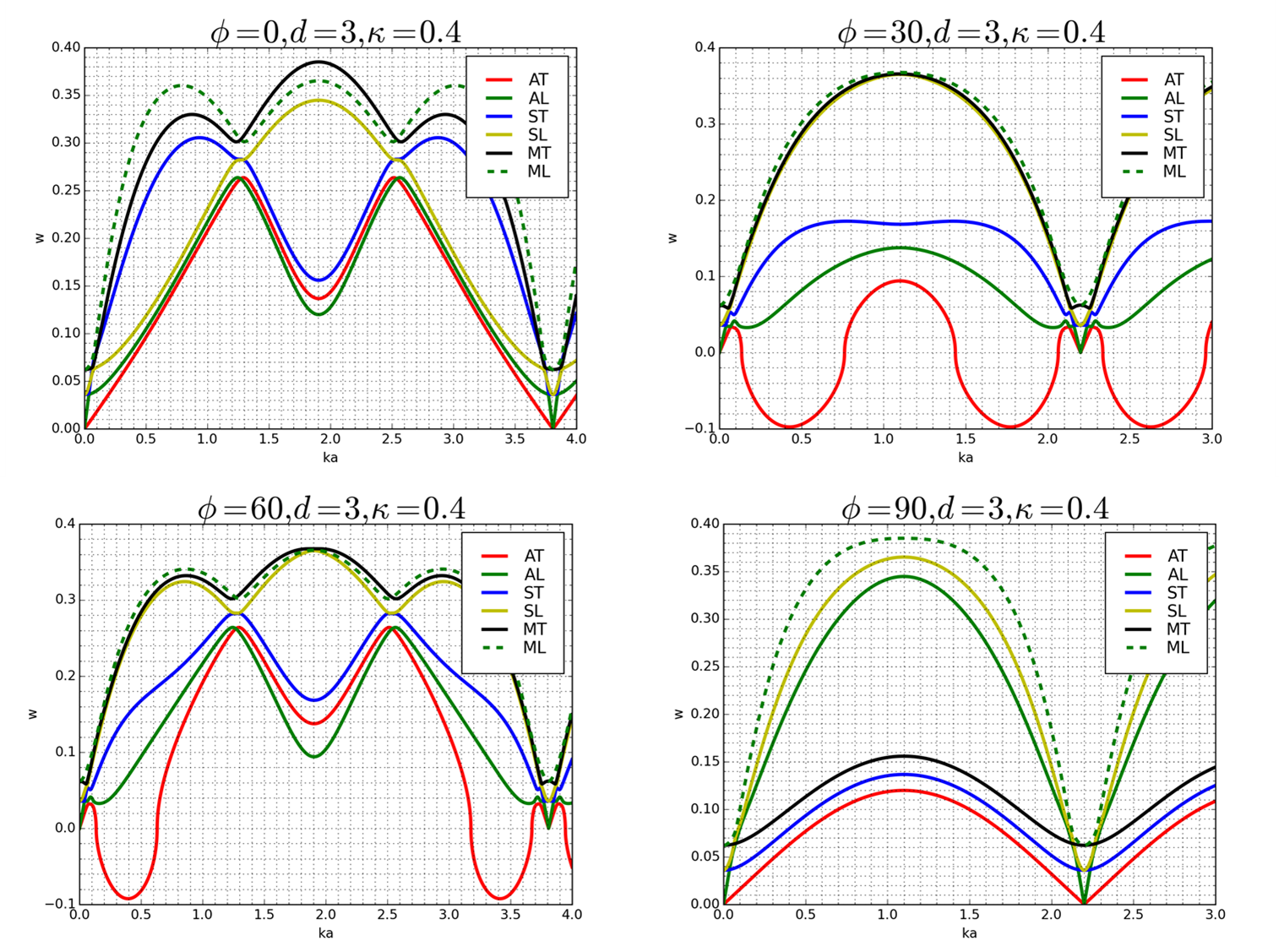}
\caption{dispersion curves for $d=3.0a$}
\label{fig:dis3}
\end{center}
\end{figure}

\begin{figure}[htbp]
\begin{center}
\includegraphics[width=\columnwidth]{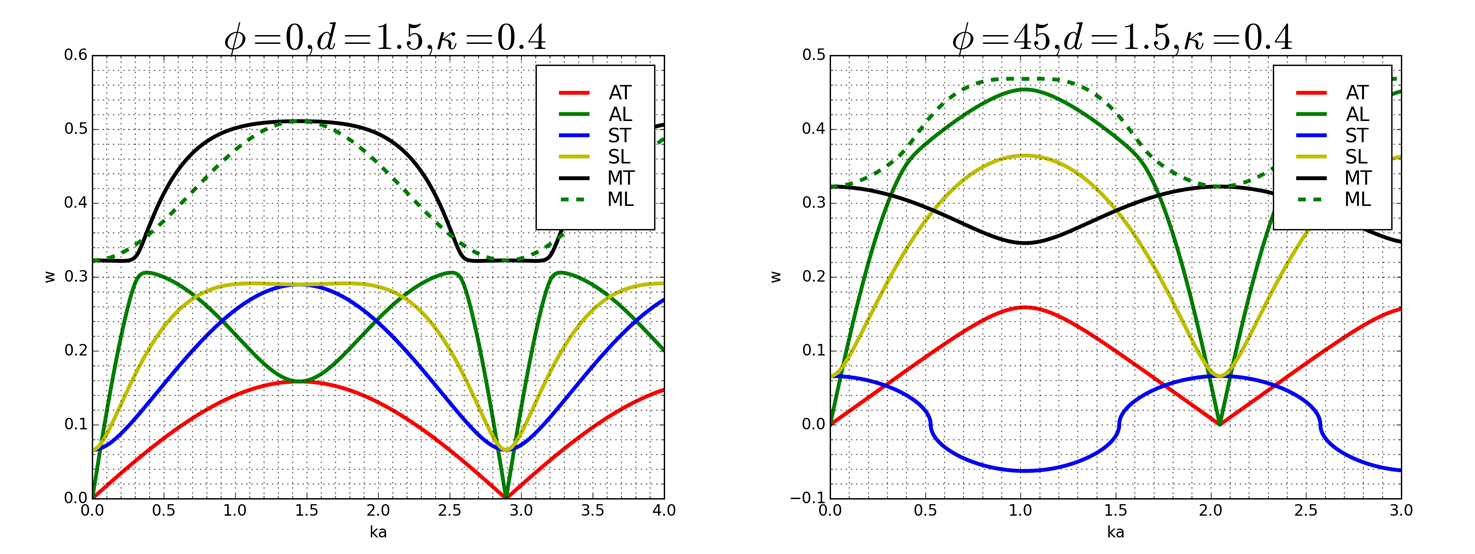}
\caption{dispersion curves for square lattice}
\label{fig:dis4}
\end{center}
\end{figure}
From Fig \ref{fig:dis2}, we can see that the dispersion is very close the that in Fig \ref{fig:dis1}, since the interlayer distance is very small. But we observed many "cross" points, where the lines with different colors form a gap.

From Fig \ref{fig:dis3}, we can see the tendency of the each graph is similar with that from a single triangle lattice, this is because the d=3, the coupling between two layers is already weak. We can see the value of one line can be negative, this means the eigenvalue of the dynamical matrix is negative, it indicates there is instability in the lattice structure.

\begin{figure}[htbp]
\begin{center}
\includegraphics[width=\columnwidth]{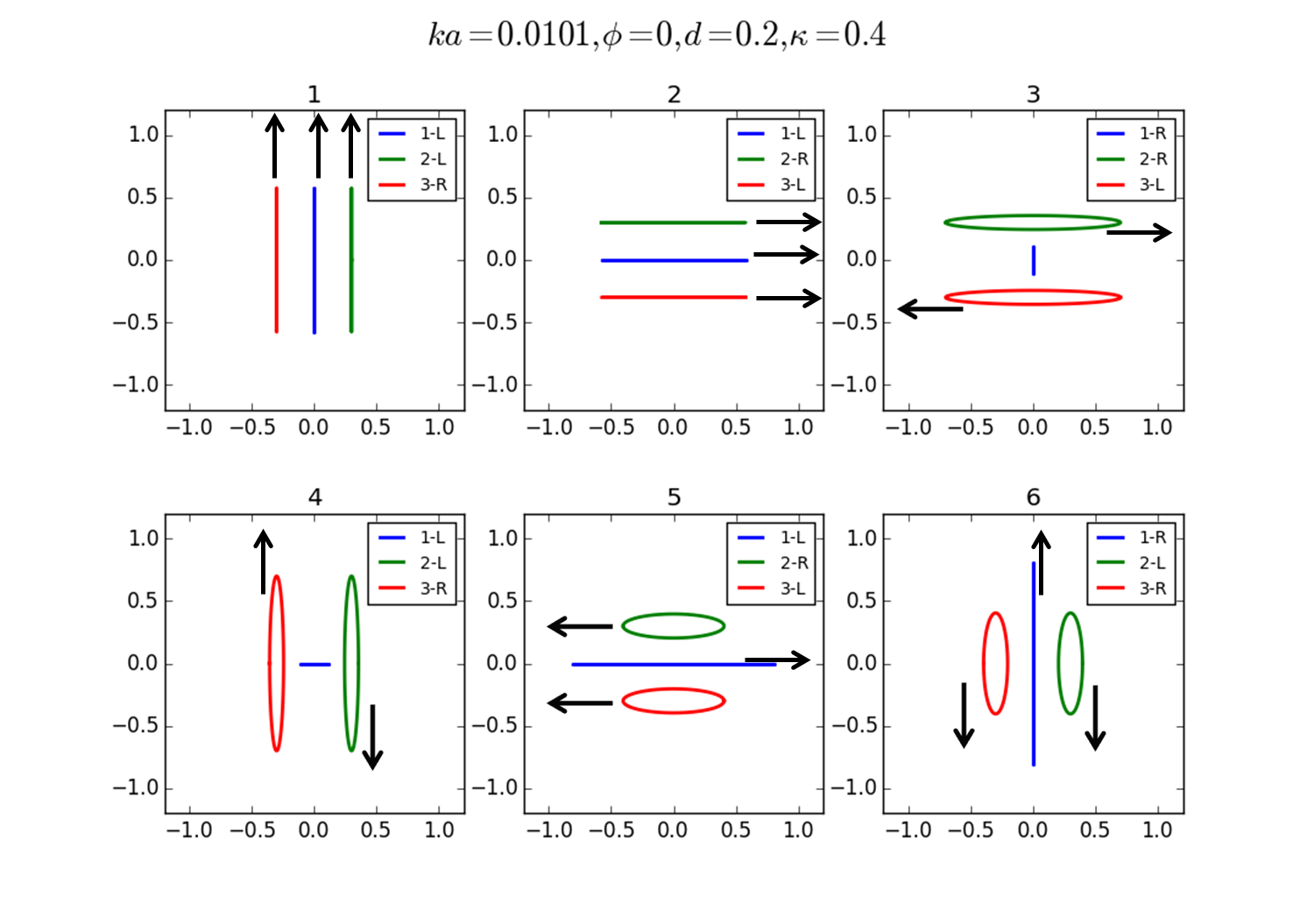}
\caption{Visualization of complex eigenvectors,1,2,3 represents middle,top, bottom layer, L,R means left and right rotation}
\label{fig:polcircle}
\end{center}
\end{figure}

\begin{figure}[htbp]
\begin{center}
\includegraphics[width=\columnwidth]{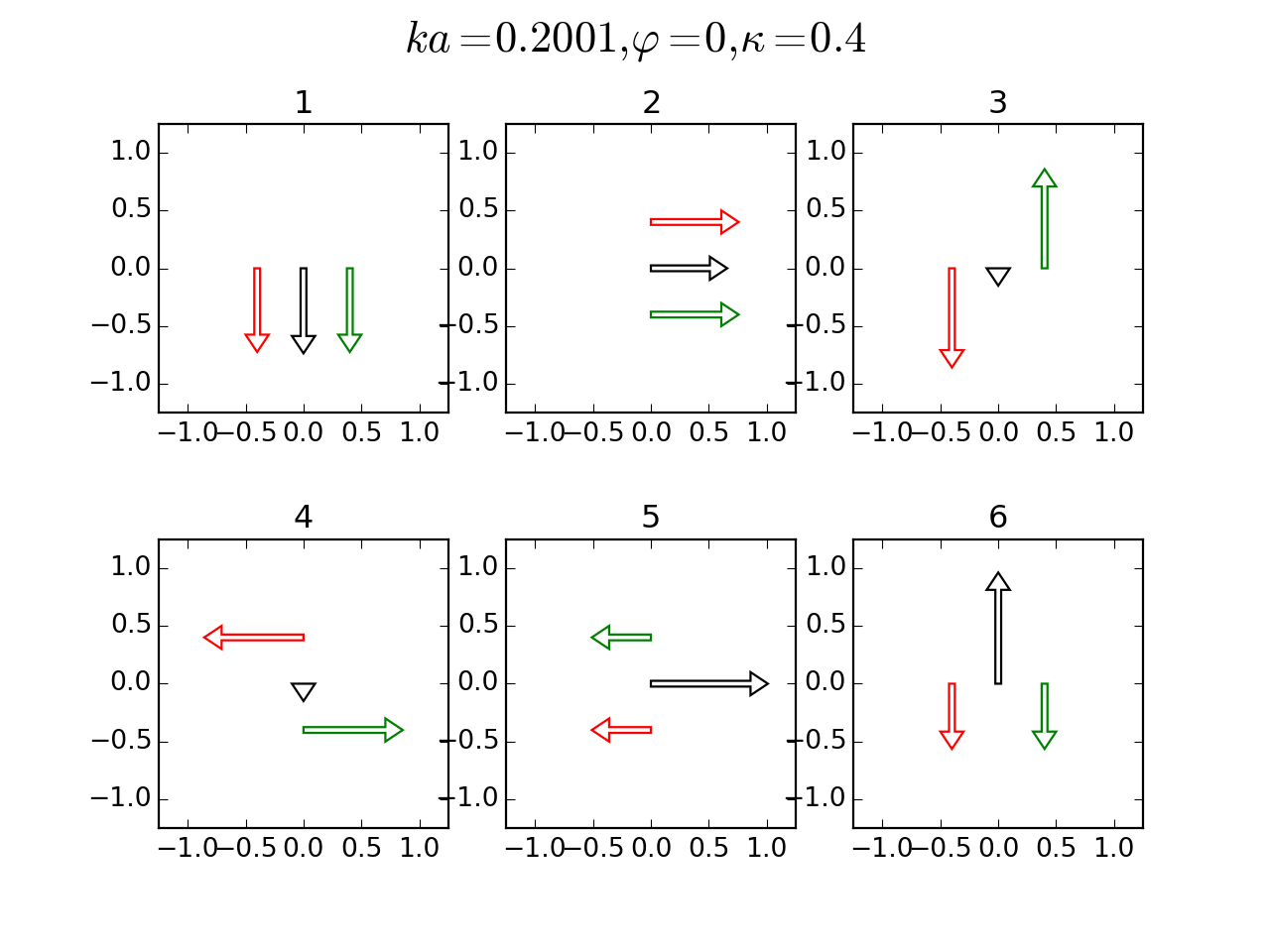}
\caption{Displacement visualization of the eigenvecotors, interlayer distance is 1.5, $\varphi$ is the propagation angle in k space respect to $k_x$ axis, the length of the arrows means its magnitude, the black, red, green corresponds to layer 1-middle,2-upper,3-lower}
\label{fig:pol1}
\end{center}
\end{figure}

From now on, we will try to analyze the pattern of the eigenvectors. When we have staggered triangular lattice, the dynamical matrix is Hermitian(some elements can have imaginary part). For overlapped rhombic or rectangle lattice, the matrix is always real symmetric. For real eigenvectors, we can simply use a 2D arrow to visualize the displacement. Suppose we get an eigenvector with six elements $(x_1,y_1,x_2,y_2,x_3,y_3)$, then we break it into three vectors $(x_1,y_1)$, $(x_2,y_2)$, $(x_3,y_3)$, we use an arrow to represent each vector.If all the tree vectors are in the same directon, it is labelled as 'A'(All), if the middle layer doesn't move, it is labelled as 'S'(Stationary), if the middle layer does move in one dirction and the other two layers are in the opposite direction, we label it as 'M'(Move). Fig \ref{fig:pol1} is the visualization for square lattice. The eigenvectors from complex matrix have complex number, it is not easy to visualize it by a simple 2D vector, so we use elliptical plot manner. We write $\frac{x_1}{y_1}$ in form of $re^\theta$, then use parametric plot of $x=rsin(wt+\theta)$, $y=sin(wt)$, the positive theta gives left rotation, the negative theta gives right rotation. Fig \ref{fig:polcircle} is an example of the lliptical plot.

Generally, whenever two dispersion curves form a gap, they exchange eigenvector type. Or in other words, The slope of the dispersion curve from a given polarization type won't change sign around a cross point. This can be seen based on the matrix structure, for example, in square lattice, all the $M_{xy}$ elements are zero.The dynamical matrix can be decomposed into two 3 by 3 sub-matrix(XX and YY). Each sub-matrix has a structure as $
\begin{pmatrix} 
A & B & B \\
B & A & C \\
B & C & A
\end{pmatrix} 
$
Its eigenvalues are $(A-C,\frac{2A+C-\sqrt{8B^2+C^2}}{2},\frac{2A+C+\sqrt{8B^2+C^2}}{2})$, the corresponding eigenvecotors are $(0,-1,1)$,$(-\frac{C+\sqrt{8B^2+C^2}}{2B},1,1)$,$(-\frac{C-\sqrt{8B^2+C^2}}{2B},1,1)$. The second and third elements in the last two eigenvectors have same value, one is 'M' mode, the other one is 'A' mode. But when B changes sign, the 'M' and 'A' mode will exchange type, where B is about zero. Since the eigenvalues difference between 'M' and 'A' mode is $\sqrt{8B^2+C^2}$, when B is zero,the difference will be a minimum.

Conclusion

We find the optimal lattice structure of a trilayer Yukawa lattice by lattice energy summation calculation. Then we calculate the in-plane collective mode dispersion. The polarization type of each mode has a exchange at each gap in the dispersion plot. Unfortunately, our dispersion result shows that the eigenvalues of the dynamical matrix is not always positive, finding the global energy minimum of the lattice is quite challenging. Future work will be done on comparing the lattice result with MD simulation result.

\begin{acknowledgments}
The authors are grateful for financial support from NSF grants PHY-1105005, the Hungarian Office for Research, Development, and Innovation NKFIH grants K-119357 and K-115805.
\end{acknowledgments}

\bibliography{references}

\end{document}